\documentclass[aps,prb,twocolumn,showpacs,preprintnumbers,amsmath,amssymb,superscriptaddress]{revtex4}%

\usepackage{graphicx}%
\usepackage{dcolumn}
\usepackage{amsmath}
\usepackage{color}

\makeatletter
\def\btt#1{\texttt{\@backslashchar#1}}%
\DeclareRobustCommand\bblash{\btt{\@backslashchar}}%
\makeatother

\begin{document}


\title{Comment on ``Muon-spin rotation studies of the superconducting properties of Mo$_3$Sb$_7$'' }
\author{R.~Khasanov}
 \email[Corresponding author: ]{rustem.khasanov@psi.ch}
 \affiliation{Laboratory for Muon Spin Spectroscopy, Paul Scherrer
Institut, CH-5232 Villigen PSI, Switzerland}
\author{P.W.~Klamut}
\affiliation{Institute of Low Temperature and Structure Research
of the Polish Academy of Sciences, Ok\'{o}lna 2, 50-422
Wroc{\l}aw, Poland }
\author{A.~Shengelaya}
 \affiliation{Physics Institute of Tbilisi State University,
Chavchavadze 3, GE-0128 Tbilisi, Georgia}
\author{I.M.~Savi\'c}
 \affiliation{Faculty of Physics, University of Belgrade, 11001
Belgrade, Serbia and Montenegro}
\author{C.~Baines}
\affiliation{Laboratory for Muon Spin Spectroscopy, Paul Scherrer
Institut, CH-5232 Villigen PSI, Switzerland}
\author{H.~Keller}
\affiliation{Physik-Institut der Universit\"{a}t Z\"{u}rich,
Winterthurerstrasse 190, CH-8057 Z\"urich, Switzerland}

\begin{abstract}
In a recent article Tran {\it et al.} [Phys.~Rev.~B {\bf 78}, 172505 (2008)] report on the result of the muon-spin rotation ($\mu$SR) measurements of Mo$_3$Sb$_7$ superconductor.  Based on the analysis of the temperature and the magnetic field dependence of the Gaussian relaxation rate $\sigma_{sc}$ they suggest that Mo$_3$Sb$_7$ is the superconductor with two isotropic $s-$wave like gaps. An additional confirmation was obtained from the specific heat data published earlier by partly the same group of authors in [Acta~Mater. {\bf 56}, 5694 (2008)]. The purpose of this Comment is to point out that from the analysis made by Tran {\it et al.} the presence of two superconducting energy gaps in Mo$_3$Sb$_7$ can not be justified. The analysis of $\mu$SR data does not account for the reduction of $\sigma_{sc}$ with increasing temperature, and, hence, yields inaccurate information on the magnetic penetration depth. The specific heat data can be
satisfactory described within the framework of the one-gap model
with the small residual specific heat component. The experimental data of Tran {\it et al.}, as well as our earlier published $\mu$SR data [Phys.~Rev.~B {\bf 78}, 014502 (2008)] all seem to be consistent with is the presence of {\it single} isotropic superconducting energy gap in Mo$_3$Sb$_7$.
\end{abstract}

\maketitle

{\it The magnetic field dependence of the $\mu$SR depolarization rate $\sigma_{sc}$}.
It is commonly accepted that the Gaussian muon-spin depolarization rate (square root of the second moment of the $\mu$SR line)  of the superconductor in the vortex state ($\sigma_{sc}$) is directly related  to the magnetic penetration depth $\lambda$ in terms of:
\begin{equation}
\sigma_{sc}=A(b)\cdot\lambda^{-2}.
 \label{eq:sigma-lambda}
\end{equation}
Here $A(b)$ is the proportionality coefficient ($b=B/B_{c2}$ is the reduced magnetic field, $B_{c2}$ is the upper critical field). One needs to stress, however, that the proportionality coefficient $A(b)$ is {\it not constant}. Its dependence on $b$ accounts for reduction of $\sigma_{sc}$ due to stronger overlapping of vortices by their cores with increasing magnetic field. As shown by Brandt,\cite{Brandt03} only for very low fields ($0.13/\kappa^2\ll b\ll 1$, $\kappa=\lambda/\xi$, $\xi$ is the coherence lengths) one can neglect the dependence of $A(b)$ on the reduced magnetic field and assume it to be constant.

The condition $A(b)=const$ is definitively not satisfied in Ref.~\onlinecite{Tran08}. The experiments were conducted for reduced fields in the range of $0.0025\leq b\leq0.025$ at $T=0.1$~K and $0.0036\leq b\leq0.036$ at $T=1$~K [$B_{c2}(0.1$~K)$\simeq2$~T and $B_{c2}(1$~K)$\simeq1.4$~T are taken from Ref.~\onlinecite{Candolfi08}]. As follows from Fig.~6 of Ref.~\onlinecite{Brandt03} at these regions $A(b)$ for Mo$_3$Sb$_7$ superconductor ($\kappa>50$) is strongly field dependent. This implies, in turn, that in order to obtain $\lambda$ from $\sigma_{sc}(B)$ data, one needs to account for reduction of $A(b)$ with increasing magnetic field.
\begin{figure}[htb]
\includegraphics[width=1.0\linewidth]{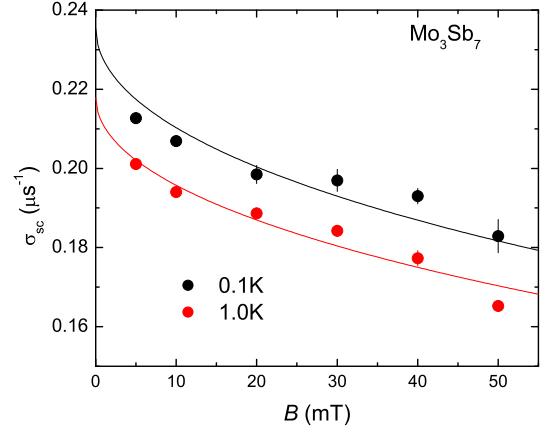}
 \vspace{-0.5cm}
\caption{(Color online) The fit of $\sigma_{sc}(B)$ dependences obtained by Tran {\it et al.}\cite{Tran08} by assuming field independent $\lambda$. See text for details.}
 \label{fig:lambda_vs_H_muSR}
\end{figure}
It may be done, {\it e.g.}, within the framework  the London model (as is made by the authors, but without accounting for some limitations of the model, see the discussion below), or by using the approach developed by Brandt in Ref.~\onlinecite{Brandt03}.

Figure~\ref{fig:lambda_vs_H_muSR} shows the results of the fit of equation:\cite{Brandt03}
\begin{eqnarray}
 \sigma_{sc}[\mu {\rm s}^{-1}]=4.83\cdot10^4\times
(1 - b) \nonumber \\
 \left[1 + 1.21\left(1 - \sqrt{b} \right)^3\right]&
\lambda^{-2}[{\rm nm}]
 \label{eq:sigma_vs_h}
\end{eqnarray}
to the experimental $\sigma_{sc}(B)$ data of Tran {\it et al.}\cite{Tran08} The Eq.~(\ref{eq:sigma_vs_h}) is derived within the framework of Ginzburg-Landau theory for the superconductor with {\it single} isotropic $s-$wave like gap.\cite{Brandt03} It describes with less than 5\% error the field variation of $\sigma_{sc}$ for an ideal triangular vortex lattice and it holds for type-II superconductors with the value of the Ginzburg-Landau parameter $\kappa\geq5$ in the range of fields $0.25/\kappa^{1.3}\lesssim b\leq1$.

An agreement of Eq.~(\ref{eq:sigma_vs_h}) with the experimental data of Tran {\it et al.}\cite{Tran08} is relatively good (see Fig.~\ref{fig:lambda_vs_H_muSR}) thus pointing to the field independent $\lambda$ and, consequently, to the presence of  only {\it one} superconducting energy gap in Mo$_3$Sb$_7$.
We want also to note that in our recent paper,\cite{Khasanov08_MoSb} which was published 3 month before the submission of Tran {\it et al.}\cite{Tran08}, $\sigma_{sc}$ as a function of magnetic field for Mo$_3$Sb$_7$ superconductor was measured up to 4 times higher field ($\mu_0H=0.2$~T, $b\simeq 0.1$) and was found to be consistent with Eq.~(\ref{eq:sigma_vs_h}) and, consequently, with the field independent magnetic penetration $\lambda$.

In reference to the interpretation of $\mu$SR data we note, that the authors of Ref.~\onlinecite{Tran08} have mixed, somehow, the statements of field dependent $\lambda$ and $\sigma_{sc}$. The muon-spin depolarization rate of the superconductor in the vortex state $\sigma_{sc}$ is {\it always} field dependent, while dependence of $\lambda$ on the magnetic field is the characteristic of unconventional superconductors (like cuprates,\cite{Sonier07,Kadono04,Khasanov09-Bi2201} pnictides,\cite{Luetkens08-Khasanov08_Sm} double gap MgB$_2$,\cite{Cubitt03_Angst04} {\it etc.}). In a single gap $s-$wave superconductor the magnetic penetration depth is found to be independent on the magnetic field.\cite{Kadono04,Khasanov05-RbOs,Khasanov06_LiPdB} In the Ref.~\onlinecite{Sonier07}, which is cited by Tran {\it et al.}\cite{Tran08} in order to justify the unconventional two-gap superconductivity in Mo$_3$Sb$_7$, Sonier refers to the field dependent penetration depth $\lambda$, but not the muon-spin depolarization rate $\sigma_{sc}$.

\vspace{0.5cm}
{\it The modified London model}.
The fit of $\sigma_{sc}$ vs. $B$ data by means of the modified
London model, used by Tran {\it et al.},\cite{Tran08} is in
favor of the ``one-gap'' picture. Note that the London model is based initially on the statement of {\it field independent} $\lambda$. By pointing to an agreement of this model with the experimental $\sigma_{sc}(B)$ data, the authors of Ref.~\onlinecite{Tran08} strongly
contradict themselves, since the  key argument of their paper is, in contrast,
the {\it field dependent} $\lambda$.

We want to stress however, that the London model uses some
simplifications and assumptions and is strictly valid for the
extreme type-II superconductor ($\lambda\gg\xi$) for fields in the
region  $0\ll B \ll B_{c2}$. The possibility to use this model in
order to describe the experimental $\mu$SR data needs to be
justified for each particular case. The authors have not done
that. On the other hand, the results of the numerical calculations
of Brand,\cite{Brandt03} which are valid for any type-II
superconductors and  in the full field region (from $0$ up to
$B_{c2}$), are free from these imperfections.

\vspace{0.5cm}

{\it Dependence of the magnetic penetration depth $\lambda$ on temperature}.
\begin{figure}[htb]
\includegraphics[width=1.0\linewidth]{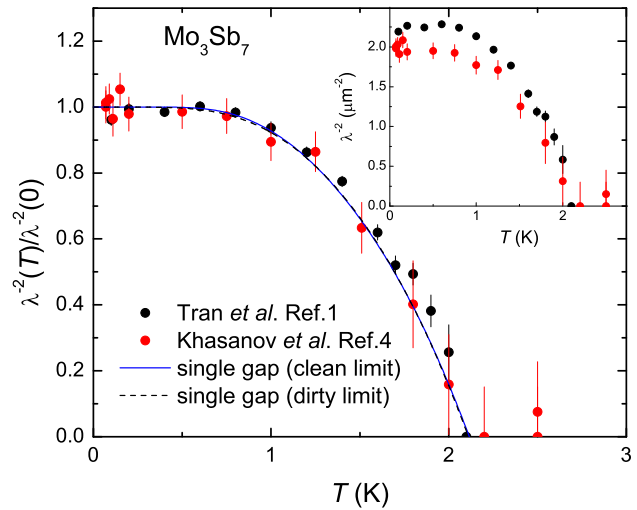}
 \vspace{-0.5cm}
\caption{(Color online) $\lambda^{-2}(T)/\lambda^{-2}(0)$ dependences of Mo$_3$Sb$_7$ reconstructed from $\sigma_{sc}(T)$ data of Tran {\it et al.}\cite{Tran08} (black circles) and that reported in Ref.~\onlinecite{Khasanov08_MoSb} (red circles). The solid lines represent the single-gap fit of $\lambda^{-2}(T)/\lambda^{-2}(0)$ data assuming Mo$_3$Sb$_7$ is the superconductor within the clean (solid curve) and the dirty (dashed curve) limit (after Ref.~\onlinecite{Khasanov08_MoSb}). The inset show the corresponding $\lambda^{-2}(T)$ dependences.  }
 \label{fig:lambda_vs_T}
\end{figure}
For the experiment conducted in constant magnetic field one needs, in addition, to account for dependence of the coefficient $A(b)$, which relates the muon-spin depolarization rate $\sigma_{sc}$ to the penetration depth $\lambda$ [see Eq.~(\ref{eq:sigma-lambda})], on temperature. It is caused by the temperature dependence of the the second critical field $B_{c2}$ and, as a consequence, that of $b=B/B_{c2}(T)$.
Obviously, this needs to be considered in order to reconstruct $\lambda(T)$ from $\sigma_{sc}(T)$ obtained experimentally. The detailed description of the reconstruction procedure (also in application to Mo$_3$Sb$_7$) is given in Refs.~\onlinecite{Khasanov08_MoSb}, \onlinecite{Khasanov06_LiPdB} and \onlinecite{Landau07_Khasanov08}.

Figure~\ref{fig:lambda_vs_T} shows $\lambda^{-2}(T)$ normalized on its value at $T=0$ for Mo$_3$Sb$_7$ superconductor. The inset represents $\lambda^{-2}(T)$ data. The solid black and the red circles refer to $\lambda^{-2}(T)$ reconstructed from $\sigma_{sc}(T)$  of Tran {\it et al.}\cite{Tran08} and that reported in Ref.~\onlinecite{Khasanov08_MoSb}, respectively. The inset of Fig.~\ref{fig:lambda_vs_T} implies that within the whole temperature region the difference between the absolute $\lambda^{-2}$ values obtained in both sets of experiments does not exceed 10\% (5\% in $\lambda$ value), which may be caused by the different sample shape (single crystalline samples in Ref.~\onlinecite{Khasanov08_MoSb} vs. fine powder in Ref.~\onlinecite{Tran08}), as well as the sample preparation procedures.

The lines in the main panel of Fig.~\ref{fig:lambda_vs_T} correspond to the fit of $\lambda^{-2}(T)$ data from Ref.~\onlinecite{Khasanov08_MoSb} by assuming that Mo$_3$Sb$_7$ is a superconductor with the {\it single} $s-$wave like energy gap within the clean (solid line) and the dirty (dashed line) limit. It is obvious that both sets of the experimental data are in agreement with the each other as well as as with the ``single-gap'' fitting curves from Ref.~\onlinecite{Khasanov08_MoSb}.

\vspace{0.5cm}
{\it The absolute value of $\lambda$}.
The 1~nm error in the absolute $\lambda$ value is unrealistic. The fit
was performed within the framework of the certain (modified
London) model, the validity of which, in application to
Mo$_3$Sb$_7$ and the conditions of the experiment, was not
justified. The fit of the $\sigma_{sc}(B)$ data by using Eq.~(\ref{eq:sigma_vs_h}) [see Fig.~\ref{fig:lambda_vs_H_muSR}] results in $\lambda(0.1$~K)$=673(3)$~nm which is 8~nm higher than $\lambda(0.1$~K)$=665(1)$~nm reported by Tran {\it et al.}\cite{Tran08} In addition, the authors did not account for any other possible sources of uncertainties as, {\it e.g.}:
i)  vortex lattice disorder;
ii) different possible symmetry of the vortex lattice (triangular vs. squared);
iii) non-gaussian line shape of the $\mu$SR line which is expected
to be seen even in a powder sample of the isotropic (weakly anisotropic)
superconductor;
iv) the background contribution from the Ag backing plate which
may be influenced by the magnetic field expelled by Mo$_3$Sb$_7$
superconductor, {\it etc.}
None of them were discussed by Tran {\it et al.} in
Ref.~\onlinecite{Tran08}. For these reasons and accounting for the
uncorrect assumption about temperature independent proportionality
between $\lambda^{-2}$ and $\sigma_{sc}$ (see the discussion above), we call the penetration depth data presented
by Tran {\it et al.}\cite{Tran08} ``inaccurate''.

\vspace{0.5cm}

{\it Temperature dependence of the electronic specific heat}.
\begin{figure}[htb]
\includegraphics[width=1.0\linewidth]{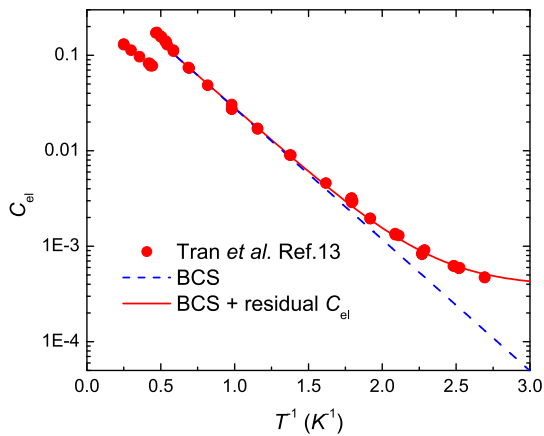}
 \vspace{-0.5cm}
\caption{(Color online) The electronic specific heat $C_{el}$ as a function of $T^{-1}$ after Ref.~\onlinecite{Tran08_Cel}. Lines correspond to the one gap fit with (the solid line) and without (dashed line) the residual electronic specific heat. Note the logarithmic $C_{el}$ scale. }
 \label{fig:cel_vs_T}
\end{figure}
One of the arguments pointing to the presence of two superconducting energy gaps in Mo$_3$Sb$_7$ was an agreement of the gap values obtained in Ref.~\onlinecite{Tran08} with that deduced by Tran {\it et al.}\cite{Tran08_Cel} in specific heat experiments ($\simeq13$\% and $\simeq5$\% difference in the absolute values of the the big and the small gap, respectively). Fig.~\ref{fig:cel_vs_T} represents the specific heat data from Ref.~\onlinecite{Tran08_Cel} together with the fits based on the ``one-gap'' BCS model. Note that, the simplest assumption about the presence of small temperature independent residual electronic specific heat, which may be easily caused by the presence of small inclusions of metallic Mo, leads to good agreement of the ``one-gap'' fit with the experimental data (see Fig.~\ref{fig:cel_vs_T}, note the logarithmic $C_{el}$ scale).

\vspace{0.5cm}

{\it Conclusions.}
The fact that the "two-gap" fits performed by Tran {\it et al.} in Refs.~\onlinecite{Tran08} and \onlinecite{Tran08_Cel}  lead to reasonable agreement between the proposed description and the experiment is obvious. Using a model with more parameters would always yield a more satisfactory fit. However, there is neither a statistical nor a physical justification for introducing more than one gap parameter in the description. The ``one-gap'' model provide already a statistically sound fit to the $\mu$SR as well as specific heat data.

In our opinion, the presence of two superconducting energy gaps in Mo$_3$Sb$_7$ may not find its justification in the experimental data presented by Tran {\it et al.} in Refs.~\onlinecite{Tran08} and \onlinecite{Tran08_Cel}. The field dependence of the muon-spin depolarization rate $\sigma_{sc}$ is well described by assuming the field independent magnetic penetration depth $\lambda$. The temperature dependences of $\lambda^{-2}$ and the electronic specific heat are consistent with what is expected for a BCS superconductor with the single $s-$wave like energy gap.

\end{document}